\documentclass[prb,aps,showpacs,prb,amsmath,floatfix,twocolumn]{revtex4}
\usepackage[]{graphicx}
\usepackage{graphicx}% Include figure files
\usepackage{dcolumn}% Align table columns on decimal point
\usepackage{bm}% bold math

\begin{document}
\title{Temporal evolution of Seebeck coefficient 
in an ac driven strongly correlated quantum dot}
\author{A. Goker$^{1}$ and E. Gedik$^{2}$}

\affiliation{$^1$
Department of Physics, \\
Bilecik University, \\
11210, G$\ddot{u}$l$\ddot{u}$mbe, Bilecik, Turkey
}

\affiliation{$^2$
Department of Physics, \\
Eskisehir Osmangazi University, \\
26480, Meselik, Eskisehir, Turkey
}

\date{\today}

\begin{abstract}
We study the response of the thermopower of a 
quantum dot in the Kondo regime to sinusoidal 
displacement of the dot energy level via a gate 
voltage using time dependent non-crossing approximation 
and linear response Onsager relations. Instantaneous 
thermopower begins to exhibit complex fluctuations when the
driving amplitude is increased at constant driving
frequency. We also find that the time averaged 
thermopower decreases steadily until it saturates 
at constant driving amplitude as a function of 
inverse driving frequency. On the other hand, time 
averaged thermopower is found to be quite sensitive
to ambient temperature at all driving frequencies
for large driving amplitudes. We discuss the underlying 
microscopic mechanism for these peculiarities based 
on the behaviour of the dot density of states.
\end{abstract}



\thispagestyle{headings}

\maketitle

\section{Introduction}

Single electron transistors consisting of
a nanodevice placed between electrodes have received
much attention recently since they are believed
to carry great promise as a future replacement
for conventional MOSFET which will reach 
its limit at 10 nm in 2016 \cite{Likharev03}. 
Unparalleled precision enabled by nanotechnology 
revolution brought immense momentum to this field. 
Determination of the switching behaviour of these 
devices in the presence of strong electron-electron
correlations is a requirement for their utilization
in electrical circuits. Moreover, investigation 
of transient electron dynamics \cite{LuetAl03Nature} 
should also help fine tuning of quantum computers
\cite{ElzermanetAl04Nature} as well as
single electron guns \cite{FeveetAl07Science}.
    
Initial investigations on the effect of Kondo 
correlations on the transient current flowing
immediately after after abrupt changing 
of the gate or bias voltage \cite{NordlanderetAl99PRL,
PlihaletAl00PRB,MerinoMarston04PRB} uncovered different 
time scales \cite{PlihaletAl05PRB,IzmaylovetAl06JPCM}. 
Since it takes Kondo resonance considerable amount 
of time to get fully formed \cite{NordlanderetAl99PRL},
strong interference between the Kondo resonance 
and the Van Hove singularities in the contacts' 
density of states results in damped oscillations 
in the long timescale for an asymmetrically coupled 
system \cite{GokeretAl07JPCM}. Better agreement 
between the experiments and theoretical predictions
can be attained by using realistic density 
of states for the contacts with the help of 
ab initio methods \cite{GokeretAl10PRB,GokeretAl11CPL}.

Transport experiments can derive additional benefit
from the measurement of thermopower (Seebeck coefficient) 
$S$ since its sign is a benchmark providing direct 
information about the alignment of the impurity orbitals 
with respect to the Fermi level of the contacts. 
Consequently, this provides unambiguous evidence about 
the existence of Kondo resonance at the Fermi level of 
the contacts. Recent experiments measuring thermoelectricity 
in molecular junctions yielded key signatures confirming 
these predictions \cite{Reddyetal07Science,Bahetietal08NL,Malenetal09NL,TangetAl10APL}.

Previous theoretical studies on this set-up mainly
focused on steady-state behaviour of thermopower.
Different groups independently observed that the 
position of the dot level is critical to determine 
the sign of thermopower in Kondo regime
\cite{DongetAl02JPCM,CostietAl10PRB}. In the first 
attempt to study the transient behaviour of thermopower 
\cite{Goker2012}, it was uncovered that the inverse of 
the saturated decay time of thermopower to its steady 
state value is equal to the Kondo temperature when 
the dot level is abruptly moved to a position close 
to the Fermi level such that the Kondo resonance is 
present in the final position.

It is quite interesting to investigate the effect of 
changing the left and right tunnel couplings of 
the dot to the electrodes sinusoidally with a phase lag
\cite{ArracheaetAl08PRB}. If the phase lag is nonzero, 
this results in asymmetrical coupling. Another interesting
situation arises when the dot energy level is subjected to 
time dependent ac perturbations in symmetrical coupling. 
This enables to monitor the evolution of thermopower in 
real time when the Kondo temperature changes constantly as a 
result of the oscillatory motion of the dot energy level 
between the leads \cite{AlhassaniehetAl05PRL,KiselevetAl06PRB}. 
This type Kondo shuttling has been made possible with 
experimental advances on controlling dot-lead coupling 
\cite{ScheibleetAl04PRL,ParksetAl07PRL}. We previously 
studied the behaviour of current in this set-up and 
found that the time averaged conductance exhibits 
considerable deviation from its monotonic decrease 
as a function of applied bias when the bias is equal 
to the driving frequency \cite{Goker08SSC}. It is our 
aim in this paper to study both the instantaneous and 
time averaged values of the thermopower when the dot 
level is driven sinusoidally by means of a gate voltage.
  
\section{Theory}  

The physical set-up of a single electron transistor consists 
of a single discrete degenerate energy level $\varepsilon_{dot}$ 
coupled to continuum electrons in the electrodes. This is a 
quantum impurity problem and it can be represented by the 
Anderson Hamiltonian
\begin{eqnarray}
H(t)&=&\sum_{\sigma} \varepsilon_{dot}(t)n_{\sigma}
+\sum_{k\alpha\sigma}\varepsilon_{k}n_{k\alpha\sigma}
+{\textstyle\frac{1}{2}}{\sum} U_{\sigma,\sigma'}n_{\sigma}n_{\sigma'}+ \nonumber \\
& &\sum_{k\alpha\sigma} \left[ V_{\alpha}(\varepsilon_{k\alpha},t)c_{k\alpha\sigma}^{\dag}c_{\sigma}+
{\rm H.c.} \right],
\label{Anderson}
\end{eqnarray}
adequately, where $c^\dagger_\sigma$ ($c_\sigma$) and 
$c^\dagger_{k\alpha\sigma}$ ($c_{k\alpha\sigma}$) with 
$\alpha$=L,R create (annihilate) an electron of spin 
$\sigma$ in the dot energy level and in the left(L) and 
right(R) electrodes respectively. The $n_\sigma$ and 
$n_{k\alpha\sigma}$ are the number operators for the dot 
level and the electrode $\alpha$. $V_{\alpha}$ are the 
tunneling amplitudes between the electrode $\alpha$ and 
the quantum dot. The Coulomb repulsion energy $U$ is 
assumed to be infinity hence the occupation of 
the dot level is strictly restricted to unity. In this paper,
we will adopt atomic units where $\hbar=k_{\rm B}=e=1$.

In our procedure, we perform  slave boson transformation 
for the Anderson Hamiltonian. The electron operators on 
the dot are rewritten in terms of a massless (slave) boson 
operator and a pseudofermion operator as
\begin{equation}
c_{\sigma}=b^{\dagger} f_{\sigma},
\end{equation}
subject to the requirement that 
\begin{equation}
Q=b^{\dagger}b+\sum_{\sigma}f^{\dagger}_{\sigma}f_{\sigma}=1.
\end{equation}
The last requirement is essentially a bookkeeping tool that
prevents the double occupancy in the original Hamiltonian
when $U \rightarrow \infty$. The main advantage of using this 
transformation is that we can now safely drop the quartic 
Hubbard term since its effects are now incorporated in other 
terms anyway. The transformed Hamiltonian then becomes
\begin{eqnarray}
H(t)&=&\sum_{\sigma}\epsilon_{dot}(t)n_{\sigma}+ \nonumber \\
& &\sum_{k\alpha\sigma}\left [\epsilon_{k}n_{k\alpha\sigma}+
V_{\alpha}(\varepsilon_{k\alpha},t)c_{k\alpha\sigma}^{\dag}
b^{\dag}f_{\sigma}+{\rm H.c.} \right],
\end{eqnarray}
where $f_{\sigma}^{\dag}(f_{\sigma})$ and $b^{dag}(b)$
with $\alpha$=L,R create(annihilate) an electron of spin 
$\sigma$ and a slave boson on the dot respectively. 

Assuming no explicit time dependency for the hopping 
matrix elements, the coupling of the quantum dot to 
the electrodes for a symmetrically coupled system 
can be expressed as $\Gamma(\epsilon)=\bar{\Gamma} \rho(\epsilon)$,
where $\bar{\Gamma}$ is a constant given by
$\bar{\Gamma}=2\pi|V(\epsilon_f)|^2$ and $\rho(\epsilon)$
is the density of states function of the electrodes.

The retarded Green function can now be rewritten
in terms of the slave boson and pseudofermion
Green functions \cite{GokeretAl07JPCM} as 
\begin{eqnarray}
G^R(t,t_1) &=& -i\theta(t-t_1)[G^R_{pseudo}(t,t_1)B^<(t_1,t) \nonumber \\
& & +G^<_{pseudo}(t,t_1)B^{R}(t_1,t)]
\end{eqnarray}

\begin{figure}[htb]
\centerline{\includegraphics[angle=0,width=9.4cm,height=6.8cm]{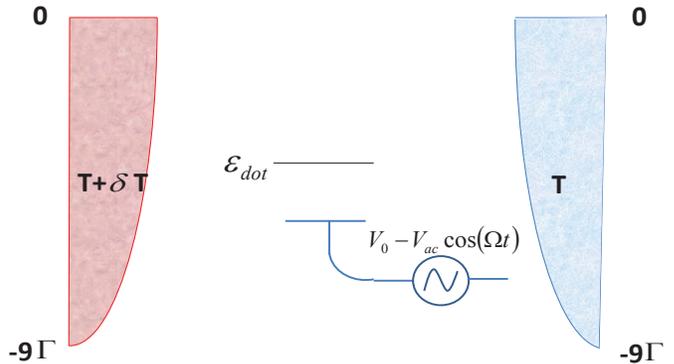}}
\caption{
This figure illustrates the density of states of both 
contacts alongside with the application of an ac gate 
voltage to the dot energy level as well as the 
temperature gradient between the contacts. 
}
\label{Schematic}
\end{figure}

It is a quite a cumbersome and highly nontrivial 
task to compute the values of these double time 
Green functions in real time as the dot level is 
in constant motion. We accomplish this by solving
coupled integro-differential Dyson equations in a 
two-dimensional cartesian grid. However, the self-energy 
of the pseudofermion and slave boson must be built 
into these equations in order to be able to obtain 
a solution. To this end, we invoke non-crossing 
approximation(NCA) to express the pseudofermion and 
slave boson self-energies \cite{ShaoetAl194PRB,IzmaylovetAl06JPCM}.
As the name implies, NCA ignores higher order corrections 
where propagators cross each other. Consequently, it 
provides erroneous results in finite magnetic fields 
and when the ambient temperature is an order of magnitude 
smaller than Kondo energy scale. We will avoid these 
cases in this paper. Main advantage of NCA is that it 
is known to give accurate results for dynamical quantities. 
Upon solving the Dyson equations properly, their values 
are stored in a square matrix which is translated one 
step at a time in diagonal direction.

In linear response, the conductance of 
the device is given by
\begin{equation}
G(t)=\frac{L_{11}(t)}{T},
\label{Seebeck}
\end{equation}
and the Seebeck coefficient or
the thermopower is expressed as
\begin{equation}
S(t)=\frac{L_{12}(t)}{T L_{11} (t)},
\label{Seebeck}
\end{equation}
where the Onsager coefficients are
\begin{eqnarray}
& & L_{11} (t)= T \times \nonumber \\
& & Im \left(\int_{-\infty} ^t dt_1 \int \frac{d\epsilon}{2\pi} e^{i\epsilon(t-t_1)} \Gamma(\epsilon) G^r (t,t_1) \frac{\partial f(\epsilon)} {\partial \epsilon}\right)
\label{Onsager1}
\end{eqnarray}
and
\begin{eqnarray}
& &L_{12} (t)= T^2 \times \nonumber \\
& & Im \left(\int_{-\infty} ^t dt_1 \int \frac{d\epsilon}{2\pi} e^{i\epsilon(t-t_1)} \Gamma(\epsilon) G^r (t,t_1) \frac{\partial f(\epsilon)} {\partial T}\right).
\label{Onsager2}
\end{eqnarray}
where $f(\epsilon)=\frac{1}{1+e^{\beta \epsilon}}$ is 
the Fermi-Dirac distribution with $\beta=\frac{1}{T}$.

Instantaneous conductance exhibits oscillations
between zero occurring at the lowest position of
the dot level and a maximum taking place at its 
closest location to the Fermi level. We previously 
analyzed the conductance for this set-up in detail 
\cite{Goker08SSC} and presented the time-averaged
density of states for the dot as well.  

\section{Results and Discussion}

In this work, we will investigate the behaviour of
Eq.~(\ref{Seebeck}) for a quantum dot in Kondo
regime whose energy level is driven sinusoidally 
via a gate voltage. We will report our results in
atomic units therefore one needs to multiply
our thermopower values with $k_B/e$ to obtain SI
value in terms of Volts/Kelvin. This conversion
factor turns out to be 0,86$\times 10^{-4}$. This 
sinusoidal motion results in periodic modulations 
of the main Kondo resonance and its satellites. 
Time averaged thermopower has been previously studied 
in this system neglecting the Kondo effect \cite{ChietAl12JPCM}. 
We would like to make this model more realistic 
by including the Kondo effect in this analysis since 
strong electron correlation effects play
a crucial role in confined nanostructures like
quantum dots. We will only consider the linear
response instantaneous thermopower due to the 
validity of Onsager relations in this regime.

Kondo effect has long been a trademark of many body 
physics taking place at sufficiently low temperatures
due to a hybridization between the net spin localized 
inside an impurity and the continuum electrons of a 
nearby metal. Its main manifestation is a sharp resonance 
around the Fermi level of the metal arising when the 
impurity level is located below the Fermi level. 
The linewidth of this Kondo resonance is roughly 
equal to a low energy scale called the Kondo temperature 
which is denoted by $T_K$ and expressed as
\begin{equation}
T_K \approx \left(\frac{D\Gamma}{4}\right)^\frac{1}{2}
\exp\left(-\frac{\pi|\epsilon_{\rm dot}|}{\Gamma}\right),
\label{tkondo}
\end{equation}
In Eq.~(\ref{tkondo}) $D$ is the half bandwidth of the 
conduction electrons while $\Gamma=\bar{\Gamma} \rho(\epsilon_f)$.

\begin{figure}[htb]
\centerline{\includegraphics[angle=0,width=8.7cm,height=6.0cm]{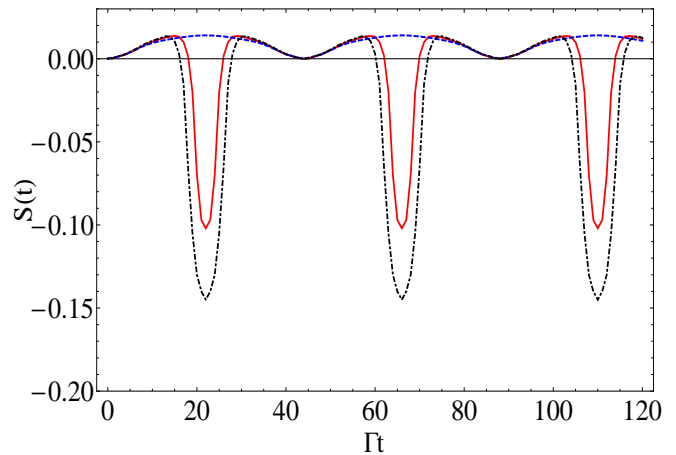}}
\caption{
This figure shows the instantaneous thermopower 
$S(t)$ immediately after the gate voltage has 
been turned on for driving amplitudes of 
$2.0\Gamma$ (blue dashed), 
$2.5\Gamma$ (red solid) and
$3.0\Gamma$ (black dot dashed) 
at $T$=0.003$\Gamma$ and
$\Omega$=0.14$\Gamma$ 
in linear response.
}
\label{Fig2}
\end{figure}

We will first consider the behaviour of the 
instantaneous value of the thermopower when 
the dot energy level is modulated sinusoidally 
via a gate voltage. We can represent the 
instantaneous behaviour of the dot level as
\begin{equation}
\epsilon_{dot}(t)=-5\Gamma-A cos(\Omega t).
\label{dotlevel}
\end{equation}
where $A$ is the driving amplitude and $\Omega$ 
is the driving frequency. The dc value of the dot energy 
level was chosen as $-5\Gamma$ to ensure that the dot
level lies sufficiently below the Fermi level such 
that the Kondo temperature is much lower than the 
ambient temperature. In principle, it is possible to 
choose a lower dot energy level as dc value. However, 
it would be problematic with our parameters because 
we would need to increase the driving amplitudes even 
higher in that case to approach the Fermi level sufficiently 
and the dot energy level would go beyond the bandwidth 
which is at $-9\Gamma$ in its lowest position. This type
of modulation induces a continuous change in Kondo 
temperature $T_K$ since its value depends on the 
dot energy level in Eq.~(\ref{tkondo}). Consequently, 
the shape of the Kondo resonance pinned to the Fermi 
level of the contacts constantly changes. The parabolic 
structure of the density of states of the contacts 
and the temperature gradient applied to them 
is depicted schematically in Fig.~\ref{Schematic}
alongside with the application of gate voltage
to the dot energy level. 
 
Instantaneous thermopower is shown in Fig.~\ref{Fig2}
for three different driving amplitudes $A$ immediately
after the gate voltage has been turned on at 
constant driving frequency and ambient temperature. 
Since the dot level is oscillating sinusoidally, 
the value of the Kondo temperature is fluctuating
alongside with it between zero and a maximum 
value. These maximum values are $T_K=0,0022\Gamma$,
$T_K=0,00055\Gamma$ and $T_K=0,00011\Gamma$  
for driving amplitudes of $3\Gamma$, $2,5\Gamma$ 
and $2\Gamma$ respectively. In this figure, 
thermopower slowly acquires a positive value as 
the dot level approaches the Fermi level for all 
driving amplitudes. For the smallest driving 
amplitude, it reaches a maximum positive value 
at its closest point to the Fermi level before 
it starts decreasing again as the dot level begins 
to go down.

However, this smooth behaviour changes abruptly
when we start increasing the driving amplitude.
When the driving amplitude becomes $2.5\Gamma$,
value of the instantaneous thermopower initially
acquires a positive value as in the previous
case, however it suddenly starts to decrease
as the dot level approaches the Fermi level and
dips into negative territory before it rebounds
and increases again as the dot level gets away
from the Fermi level. When the driving amplitude
is increased even further, instantaneous thermopower
starts decreasing earlier and more importantly,
it dips deeper into negative territory before
it rebounds as the dot level is pushed back. This
fluctuating behaviour is repeated over many cycles 
since the dot level's motion follows the gate 
voltage and is strictly sinusoidal consequently. 

\begin{figure}[htb]
\centerline{\includegraphics[angle=0,width=8.7cm,height=6.0cm]{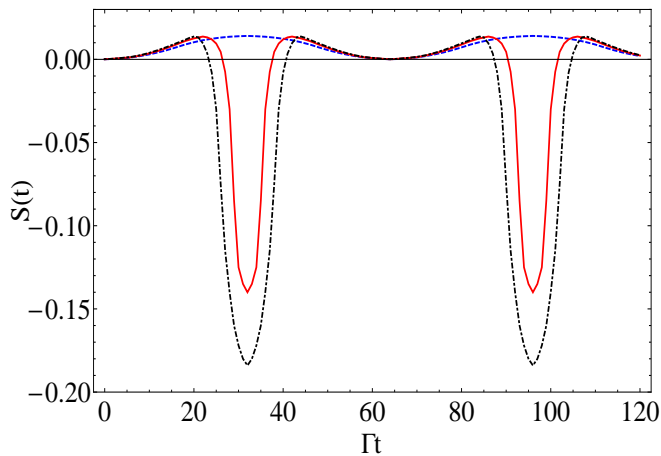}}
\caption{
This figure shows the instantaneous thermopower
$S(t)$ immediately after the gate voltage has
been turned on for driving amplitudes of
$2.0\Gamma$ (blue dashed),
$2.5\Gamma$ (red solid) and
$3.0\Gamma$ (black dot dashed)
at $T$=0.003$\Gamma$ and
$\Omega$=0.10$\Gamma$  
in linear response.
}
\label{Fig3}
\end{figure}

Fig.~\ref{Fig3} shows the instantaneous thermopower 
for the same parameters used in Fig.~\ref{Fig2}, 
but the gate voltage is applied with a smaller 
frequency this time. An obvious consequence of
this on the instantaneous thermopower is the longer 
oscillation periods for all driving amplitudes. 
Besides this trivial effect, we notice that the 
smooth oscillation behaviour for the smallest driving 
amplitude did not change much. On the other hand, 
there are significant changes for the other two larger 
driving amplitude cases compared to the larger driving 
frequency. Their fluctuation behaviour seems qualitatively 
similar but they both dive much deeper into negative 
territory as the dot level comes close to the Fermi level. 
This effect is quite pronounced for both cases.

\begin{figure}[htb]
\centerline{\includegraphics[angle=0,width=8.7cm,height=6.0cm]{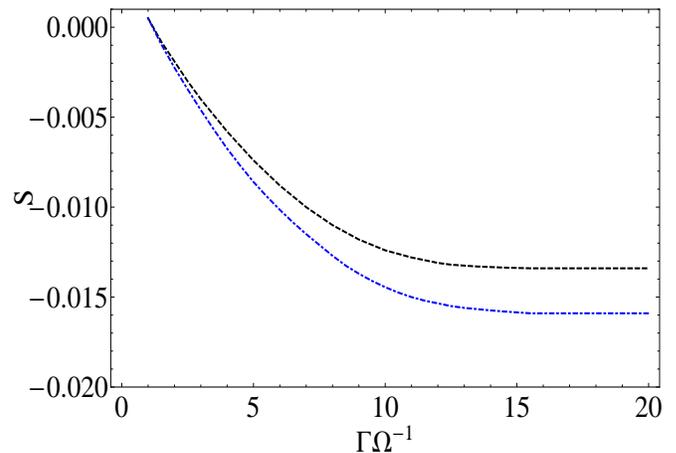}}
\caption{
This figure shows the value of the 
thermopower averaged over a period as a function
of inverse driving frequency at $T$=0.003$\Gamma$
(black dashed) and $T$=0.002$\Gamma$ (blue dot dashed)
for driving amplitude of 2.5$\Gamma$.
}
\label{Fig4}
\end{figure}

Having observed these peculiarities, one would
be naturally inclined to find out whether these
conclusions are valid for other driving
frequencies and ambient temperatures. In order
to pinpoint this issue, we calculated the
time average value of the instantaneous thermopower
over a full period for two different ambient 
temperatures at several different driving
frequencies. Time averaged thermopower for
the smallest driving amplitude was not sensitive
to the changes in temperature so we are not 
reporting it in the following. We will 
elucidate its reason later.

Fig.~\ref{Fig4} depicts the time averaged value
of the instantaneous thermopower as a function
of inverse driving frequency for two different
ambient temperatures at a constant driving amplitude
of  2.5$\Gamma$. The first conclusion that
can be drawn from this figure is that the absolute
value of the time averaged thermopower is 
always different at all driving frequencies
when the ambient temperature changes.
Moreover, absolute value of the time averaged
thermopower starts to increase for both 
ambient temperatures as the driving frequency
is decreased but it saturates at a certain 
driving frequency and stays constant thereafter.

\begin{figure}[htb]
\centerline{\includegraphics[angle=0,width=8.7cm,height=6.0cm]{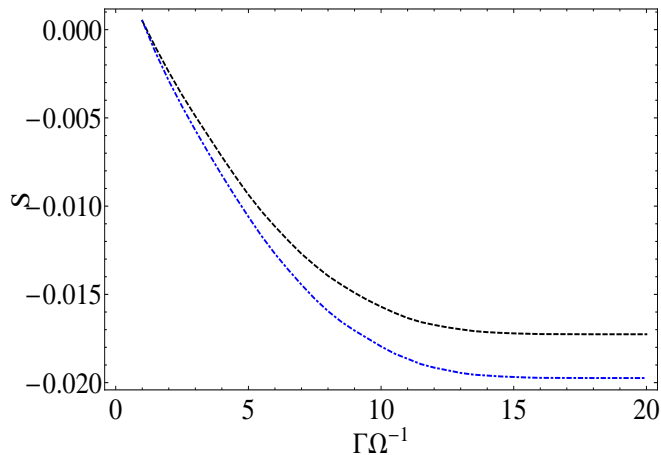}}
\caption{
This figure shows the value of the thermopower 
averaged over a period as a function of inverse 
driving frequency at $T$=0.003$\Gamma$ (black dashed)
and $T$=0.002$\Gamma$ (blue dot dashed) for driving 
amplitude of 3.0$\Gamma$.
}
\label{Fig5}
\end{figure}

This is a quite peculiar and interesting observation 
and in order to test whether it is valid for other 
driving amplitudes as well, we performed the same 
analysis for the largest driving amplitude of 3.0$\Gamma$ 
keeping all the other parameters constant. We do not
increase the driving amplitude beyond this value
to avoid entering mixed valence regime where the Kondo
and Breit-Wigner resonances overlap. We are only interested
in investigating the effect of Kondo resonance on thermopower
in this paper. The result for this driving amplitude
can be seen in Fig.~\ref{Fig5}. It is clear from this 
figure that the conclusion we had drawn is indeed valid 
here too. Namely, time averaged thermopower is quite 
sensitive to ambient temperature and it saturates when 
the driving frequency is decreased. However, there is 
an important difference. Absolute values of the time 
averaged thermopower at both ambient temperatures and 
all driving frequencies are greater than the previous 
case in Fig.~\ref{Fig4} where the driving amplitude 
was 2.5$\Gamma$. This is also an important result 
which we will dwell on below.

We now want to provide an unambiguous microscopic
explanation to the results we presented above.
It is clear that the development of the Kondo
resonance plays a crucial role in transport
of this prototypical device. Evolution of the Kondo
resonance is incorporated in the dot density
of states and the relation between the dot 
density of states and the thermopower at low
temperatures is given by the Sommerfeld 
expansion. It can be expressed as
\begin{equation}
S(T)=-\frac{\pi^2 T}{3 A(0,T)}\frac{\partial A}{\partial \epsilon} \rvert_{\epsilon=0}
\end{equation}
in atomic units. In this equation, $A(0,T)$ is 
the value of the dot density of states at Fermi level
of the contacts and $\frac{\partial A}{\partial \epsilon}$ 
is its derivative. It has been shown that the formation
of the Kondo resonance occurs slightly above the Fermi 
level \cite{CostietAl94JPCM} making the derivative of the dot 
density of states at Fermi level positive at ambient
temperatures close to or less than the Kondo temperature. 
This will give rise to a negative thermopower in this 
regime. On the other hand, Breit-Wigner resonance around 
the dot energy level is the only prominent feature of 
the dot density of states in the absence of Kondo resonance 
at ambient temperatures much greater than the Kondo temperature. 
Consequently, the thermopower would be positive if the dot 
level is positioned sufficiently away from the Fermi level 
since the dot density of states at Fermi level will lie in the 
long tail of the Breit-Wigner resonance. Thus, a shift
in the sign of the thermopower by tuning the temperature 
is a signature of the emergence of the Kondo resonance. 

When the driving amplitude is 2.0$\Gamma$, the dot 
level cannot approach the Fermi level to be able to 
develop the Kondo resonance. Therefore, T$\gg T_K$ 
for all dot levels involved in this oscillation and 
the thermopower never dips into negative territory. 
It starts from zero when the dot level is farthest 
from the Fermi level because the density of states is 
essentially zero everywhere around the Fermi level. 
However, it gradually acquires a positive value as 
the dot level approaches the Fermi level because the 
tail of the Breit-Wigner resonance starts to reach the 
Fermi level causing a negative slope there. Thermopower
reaches a maximum positive value before the dot level 
starts to retreat. The values are not sensitive to 
temperature because Kondo resonance cannot be formed
at all at any stage of this motion.

The oscillation of the dot energy level covers a larger
energy range as the driving amplitude is ramped up to 
2.5$\Gamma$. The value of the thermopower initially 
follows the value obtained when the driving amplitude 
was 2.0$\Gamma$ but it starts to deviate significantly 
when the dot level approaches the Fermi level because 
the Kondo resonance begins to emerge there. That is
accompanied with a change in slope of the density of 
states of the dot and consequently a change in sign of 
the thermopower. The thermopower sees its minimum
value when the dot level is at its closest point 
to the Fermi level because the Kondo resonance is
at its most developed form here. As the dot level
starts to retreat, the Kondo resonance is gradually
suppressed hence the thermopower becomes positive
after a while.

Similar conclusions hold for the largest driving amplitude 
of 3.0$\Gamma$ except for the fact that the instantaneous 
thermopower dips lower into negative territory since the 
dot energy level manages to get even closer to the Fermi 
level hence the Kondo resonance is more developed.

The saturation behaviour of the time averaged thermopower 
at low driving frequencies is also closely related to the 
development of the Kondo resonance. As the driving frequency 
is lowered, dot energy level begins to spend more time near 
the Fermi level. This in turn gives the Kondo resonance more 
time to develop since its full formation takes considerable 
amount of time \cite{NordlanderetAl99PRL}. Consequently, time
averaged thermopower values start to increase in negative 
territory. However, when the driving frequency is lowered 
below a certain value, we observe that the time averaged 
thermopower saturates. This is simply because the dot
level is allocated sufficient time in the vicinity of 
Fermi level anymore. 

Dependence of the time averaged thermopower curves 
to ambient temperature is quite interesting too and 
requires further elaboration. It is clear from both 
Fig.~\ref{Fig4} and Fig.~\ref{Fig5} that the difference 
between the time averaged values obtained at the same 
driving frequency increases as the driving frequency is 
reduced until it stays constant below a certain frequency.
This behaviour is further testimony to the crucial role 
development of the Kondo resonance plays in thermopower. 
At elevated driving frequencies, time averaged thermopower
is not sensitive to ambient temperature because the dot 
level cannot spend sufficient time near the Fermi level. 
Naturally, the Kondo resonance cannot be formed adequately
hence a lack of sensitivity to ambient temperature. As the 
driving frequency is reduced, the difference becomes more
pronounced and stays constant after the Kondo resonance 
gets the chance to develop itself fully below a certain 
frequency.

\section{Conclusion}

In conclusion, we studied in linear response both 
the instantaneous and time averaged values of 
thermopower for a quantum dot sinusoidally driven 
by a gate voltage. We found that the instantaneous 
thermopower exhibits complex fluctuations as the 
driving amplitude is ramped up. We interpreted the 
origin of dipping into negative territory as a sign 
of development of the Kondo resonance at Fermi level.
We studied the behaviour of the time averaged thermopower
as a function of inverse driving frequency and observed
that it tends to saturate at low driving frequencies.
Moreover, the results depend on ambient temperature
quite sensitively. We believe that the scenario we
discussed in this paper has great potential to be 
realized experimentally. The temperature gradient 
between the contacts can be achieved with laser
irradiation and the sinusoidal motion of the dot level
can easily be realized with an ac gate voltage. We 
also think that state of the art ultrafast pump-probe 
techniques \cite{Teradaetal10JPCM,TeradaetAl10Nature}
could make recording thermopower in real time
possible. Therefore, we hope to create motivation
and experimental progress in this field with
our novel results.

\section{Acknowledgments}

The authors thank T$\ddot{u}$bitak for
generous financial support via grant 111T303.

\bibliographystyle{iopams}
%\bibliographystyle{prstyx}   %>>>> makes bibtex use spiebib.bst
%\bibliography{/home/agoker/kondo08/gen} 
%\bibliography{gen}

\providecommand{\newblock}{}

\end{document}